\documentclass[journal,twoside,web]{ieeecolor}
\usepackage[colorlinks=true]{hyperref}
\usepackage[utf8]{inputenc}
\usepackage[T1]{fontenc}
\usepackage{caption}
\usepackage{subcaption}
\usepackage{todonotes}
\usepackage{amsmath, bm}
\usepackage{nccmath}
\usepackage{multirow}
\usepackage{indentfirst}
\usepackage{array}
\usepackage{varioref}
\usepackage{fancyhdr}
\usepackage{placeins}
\usepackage{threeparttable}
\usepackage{balance}
\usepackage{float}
\usepackage{generic}
\usepackage{textcomp}

\title{Deep Neural Networks Generalization and Fine-Tuning for 12-lead ECG Classification}

\author{Aram Avetisyan, Shahane Tigranyan, Ariana Asatryan, Olga Mashkova, Sergey Skorik, Vladislav Ananev, and Yury Markin
\thanks{This work was supported by the Ministry of Science and Higher Education of the Russian Federation Grant 075-15-2022-294. (Corresponding author: Aram Avetisyan)}
\thanks{Aram Avetisyan, Olga Mashkova, Sergey Skorik and Yury Markin are with the
Ivannikov Institute for System Programming of the Russian Academy of Sciences, 109004 Moscow, Russia (e-mail: a.a.avetisyan@gmail.com; o.mashkova98@gmail.com; skorik.sn99@gmail.com; ustas@ispras.ru). }
\thanks{Shahane Tigranyan and Ariana Asatryan are with the Russian-Armenian University, 375051 Yerevan, Armenia (e-mail: shahane.tigranyan99@gmail.com; arianasatryan@gmail.com). }
\thanks{Vladislav Ananev is with the Novgorod State University, 173003 Veliky Novgorod, Russia (e-mail: survial53@gmail.com). }}

\begin{document}

\maketitle
\begin{abstract}

Numerous studies are aimed at diagnosing heart diseases based on 12-lead electrocardiographic (ECG) records using deep learning methods. These studies usually use specific datasets that differ in size and parameters, such as patient metadata, number of doctors annotating ECGs, types of devices for ECG recording, data preprocessing techniques, etc.
It is well-known that high-quality deep neural networks trained on one ECG dataset do not necessarily perform well on another dataset or clinical settings.
In this paper, we propose a methodology to improve the quality of heart disease prediction regardless of the dataset by training neural networks on a variety of datasets with further fine-tuning for the specific dataset. To show its applicability, we train different neural networks on a large private dataset TIS containing various ECG records from multiple hospitals and on a relatively small public dataset PTB-XL. We demonstrate that training the networks on a large dataset and fine-tuning it on a small dataset from another source outperforms the networks trained only on one small dataset. We also show how the ability of a deep neural networks to generalize allows to improve classification quality of more diseases.

\end{abstract}

\begin{IEEEkeywords}
Electrocardiography, time series analysis, deep learning, ECG classification, fine-tuning, pre-trained neural networks
\end{IEEEkeywords}

\thispagestyle{empty}

\section{Introduction}

\IEEEPARstart{C}{ardiovascular} diseases nowadays represent the leading cause of death. Several tests are commonly used to diagnose heart conditions. For instance, tests such as echocardiograms can more accurately determine the presence of disease than others. However, some of them are generally not easily accessible to patients, require specialized equipment, and are also time-consuming and expensive.

Electrocardiography (ECG) is a non-invasive tool to assess the general cardiac condition of a patient. Due to it being a fast, painless, efficient, and low-maintenance test, it is used as the primary method for diagnosing cardiovascular diseases. In this paper, we focus on the improvement of the most common type of ECG records used by medical institutions, short 12-lead ECGs.

Despite the wide application of ECG recording, ECG analysis is a complex task that requires the expertise of a specialist with broad specific knowledge. Often not only the health but also the life of the patient depends on the timely decoding of all the data. This is further complicated by the difficulty of manual ECG analysis, which increases the likelihood of errors or incompleteness of diagnosis in interpretation. Therefore, numerous studies aim to detect abnormalities using automated methods such as applying DNNs in ECG analysis.

Automated interpretation in ECG analysis can transform the ECG into a screening tool and predictor of diseases. However, despite the great development of various deep learning methods applied to ECGs, there are still significant limitations that do not allow us to assert the success of these methods in a clinical setting with more non-unified and noisy data.

The number of digital ECGs is increasing dramatically. There are promising studies that use classical machine learning methods to detect cardiac abnormalities~\cite{alickovic2016medical,dohare2018detection}. Nevertheless, the most popular methods for analyzing ECGs are DNNs. To detect an abnormality, the researchers either use raw or preprocessed ECG signals~\cite{smigiel2021ecg,baek2021new} or convert signals to images using wavelet or short-time Fourier transform~\cite{huang2019ecg,tadesse2019cardiovascular,wang2021automatic}. Multiple surveys describe deep learning methods for ECG classification~\cite{hong2020opportunities,murat2020application}.

Most of the studies use partial and end-to-end deep learning techniques with convolutional layers and residual connections. Several approaches have already shown cardiologist-level performance for some cardiac abnormalities~\cite{hannun2019cardiologist}.

One of the main problems that arise in medical domain studies is the small amount of data. The data is often difficult to obtain because of privacy and security reasons. Consequently, few papers use large amounts of ECG data~\cite{ribeiro2020automatic,weimann2021transfer}. Even with a large dataset available~\cite{ribeiro2020automatic}, there may be challenges because the diagnosis is usually presented in text form. In this case, labels have to be extracted from textual reports, which can lead to additional errors~\cite{wang2012extracting}.

Most of the studies that analyze 12-lead ECGs use open datasets, such as PTB-XL, CPSC2018, Chapman-Shaoxing Database~\cite{wagner2020ptb,liu2018open,zheng202012}, for evaluation. These datasets consist of several thousand ECGs with fixed parameters such as duration and frequency. In addition, each dataset is preprocessed and annotated by a small group of cardiologists, who can make similar mistakes during annotation. Due to the small number of samples in open datasets, neural networks are limited to the number of abnormalities that they can classify well. Thus, DNNs trained on these datasets to predict widespread abnormalities do not show the same quality on datasets from different domains. 

Several studies demonstrate that the same DNNs trained on different datasets show significantly different results on their respective test datasets even for the most popular heart diseases such as atrial fibrillation~\cite{jo2021explainable,smisek2020cardiac,nonaka2021depth}. Such a difference in prediction performance for various ECG datasets makes the networks less robust. This makes it impossible to widely use them on clinical datasets nowadays.

In this paper, we propose a methodology to improve the accuracy of ECG classification. We demonstrate that training DNN on a large dataset assembled from a variety of different datasets and then fine-tuning it on a relatively smaller dataset improves abnormality prediction for this specific dataset. Additionally, we demonstrate that DNNs trained on a large dataset have good generalization ability for ECG analysis and can help to improve prediction results for different datasets. 

We conduct the experiments using a large, assembled dataset TIS of raw 12-lead ECG records and one of the most popular public datasets PTB-XL~\cite{wagner2020ptb}. We compare DNNs with different architectures trained on TIS and PTB-XL for 7 selected abnormalities and show that neural networks trained on TIS show stable prediction quality regardless of the test data whereas networks trained on PTB-XL start degrading on the data from a different source. We also demonstrate, for two selected abnormalities, that trained DNNs deliver results comparable to cardiologists' one.

\section{Methods}\label{sec:methods}

\subsection{Dataset retrieval and preprocessing} 
\begin{figure*}[!t]
\centering
\includegraphics[width=0.8\textwidth]{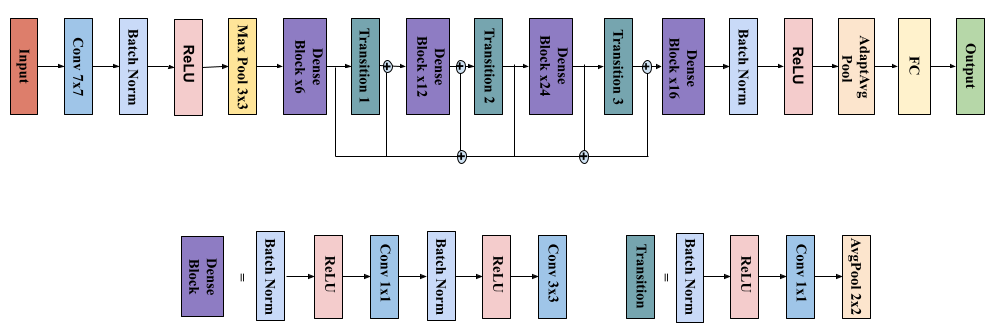}
\caption{The DenseNet architecture for ECG classification.}
\label{fig:Densenet}
\end{figure*}

Two datasets are used for analysis. The first one is a publicly available PTB-XL dataset of 21,837 ECG records~\cite{wagner2020ptb}. All ECGs in PTB-XL are 10 seconds in duration and have a 500 Hz frequency. The ECGs are preprocessed by a bandpass filter to reduce the noise and annotated by up to 2 cardiologists. The second one is the TIS dataset of more than 1,500,000 12-lead ECGs collected from multiple hospitals over 2 years with almost 57\% of labeled records annotated by more than 130 cardiologists. The data is stored in signal format and does not have any initial preprocessing. These hospitals use different devices to record ECGs in a clinical setting. As a result, the dataset is heterogeneous in various parameters: types of recording devices, frequency, duration, etc. For this research, we use 549,279 records from the TIS dataset, which are collected and annotated by the doctors for the period from December 2019 to December 2021. Ethical approval was granted by the Ethics Committee of the I.M. Sechenov First Moscow State Medical University (Protocol No. 06-23).

One of the important advantages of records in TIS is the initial annotation of the dataset with labels apart from textual diagnoses. Existing research suggests methods to extract diseases from textual diagnoses~\cite{ribeiro2020automatic,naseem2021comparative}. However, these methods perform with a certain percentage of errors, which can affect the prediction. TIS dataset is developed in a way that each doctor can select the abnormality from the list of heart diseases along with writing a diagnosis. Therefore, there is no need to use the algorithms to retrieve the labels from texts, potentially creating additional errors in the dataset.

To conduct valid experiments, TIS and PTB-XL formats should be standardized. Therefore, annotated TIS records are filtered and converted to the same data format as PTB-XL. For training and validation, we selected ECG records of 9-10 seconds duration and 500, 1000 Hz frequency. All records are resampled to 500 Hz and zero-padded to 10 seconds. For the TIS test dataset, we use only 10-second ECG records with 500 Hz frequency. We also considered only patients older than 18 years.

Errors can appear while recording ECGs in hospitals, for example, in case of a bad electrode cable connection or improperly working cardiographs. To avoid these errors, the records that had at least one lead with the constant value are removed during data preprocessing.

The output ECG signal characteristics can vary in datasets because they depend on the recording devices, the systems, and the formats in which they are stored. For example, PTB-XL and TIS records have 1.6 and 2906 average amplitude respectively, therefore, can be interpreted incorrectly by the DNNs without preprocessing. To achieve a similar average amplitude, we apply z-score normalization to both datasets after the initial transformations.

\subsection{Model architecture and training}

We use a popular DenseNet model~\cite{huang2017densely} from the family of CNNs widely used in ECG classification studies~\cite{somani2021deep} which is adapted for unidimensional data. We also reviewed the DNNs proposed in~\cite{strodthoff2020deep}. To avoid adjusting architecture to a specific dataset, we use default values for the kernel size, convolutional layers number, and node number per layer for DenseNet. We use default parameters for the other observed DNNs for the same reasons.

The observed DenseNet model consists of a Convolutional (Conv) layer with kernel size 7, followed by a max pooling layer, 4 Dense blocks, and 3 Transition blocks. The output of the last Dense block is fed into the Adaptive Average Pooling layer and the fully connected layer for classification. The first Conv layer and each layer inside the Dense block are followed by batch normalization and ReLU activation function. For the Transition block, we use average pooling with kernel size 2. The architecture of DenseNet is presented in Fig.~\ref{fig:Densenet}.

We split both TIS and PTB-XL datasets into train and test parts and use the test dataset to evaluate the DNNs. The test datasets are fixed for the experiments and the models trained on TIS and PTB-XL are both evaluated on the same data. After filtering, preprocessing, and splitting the data 549,279 ECG records from the TIS dataset are selected for training and validation, and 31,872 records are used to evaluate the DNNs. The characteristics of TIS are presented in Table~\ref{table:tis}. ECGs recorded during two selected months are used from the TIS dataset for tests. PTB-XL has a parameter in a database that helps to split the data into 10 stratified folds. We use this parameter to divide the dataset for training and testing. Hence, ECG records of the 10th fold are selected for the test dataset, other records are used for training and validation.

For each selected abnormality, we trained a binary classifier. Each trained neural network has the same input format of the ECG records of 10 seconds duration with a 500 Hz frequency. We use binary cross-entropy with logit loss which is minimized with the Adam optimizer~\cite{kingma2014adam}. The optimizer has default parameters with a learning rate of 0.003. Due to class imbalance, we use weighted loss, giving higher weight to the positive class based on the proportion of presence. The learning rate is reduced by a factor of 0.8 whenever the validation loss does not present any improvement for 3 epochs in a row. The convolutional layers' weights are initialized with Kaiming normal values~\cite{he2015delving}. The rest of the layers are initialized with constant weight and zero bias. The training runs for 100 epochs with an early stopping with the patience of 20. Finally, the DNN with the best validation loss during training is selected.

\begin{table}[b]
\caption{TIS dataset patient and ECG Characteristics}
\label{table:tis}
\centering
\begin{tabular}{|c|c|c|}
\hline
& \textbf{Train+Val} & \textbf{Test} \\ \hline
\textbf{Total} & 549,279 & 31,872 \\ \hline
\textbf{Age} & & \\ 
\textit{18-29} & 51,755 (9.4\%) & 3,704 (11.6\%) \\ 
\textit{30-49} & 133,255 (24.3\%) & 8,374 (26.3\%) \\ 
\textit{50-69} & 260,917 (47.5\%) & 15,024 (47.1\%) \\ 
\textit{>= 70} & 103,352 (18.8\%) & 4,770 (15.0\%) \\ \hline
\textbf{Sex} & & \\ 
\textit{male} & 199,125 (36.3\%) & 12,222 (38.3\%) \\ 
\textit{female} & 350,154 (63.7\%) & 19,650 (61.7\%) \\ \hline
\textbf{Frequency} & & \\ 
\textit{500 Hz} & 432,925 (78.8\%) & 31,872 (100.0\%) \\
\textit{1000 Hz} & 116,354 (21.2\%) & 0 (0.0\%) \\ \hline
\textbf{Number of doctors} & 129 & 79 \\ \hline
\end{tabular}

\end{table}

\subsection{Hyperparameter tuning}

We use stratified cross-validation with 5 folds to find the best values for hyperparameters for both datasets. We repeated the selection, training, and evaluation process 15 times. To select optimal values for learning rate and reducing factor, we use the following options: learning rate in the range of 0.001 to 0.003 and factor between 0.4 to 0.8. The best metrics are achieved for the learning rate equal to 0.003 and the factor to 0.8. On the one hand, such a choice is enough to ensure convergence with little steps, on the other hand, it is small enough to avoid divergent behavior. Activation function ReLU is chosen among ReLU, ELU, SELU, and LeakyReLU. For the batch size, we selected 256, the maximum value that can fit in the memory of the GPU.

\begin{figure}
\includegraphics[width=\columnwidth]{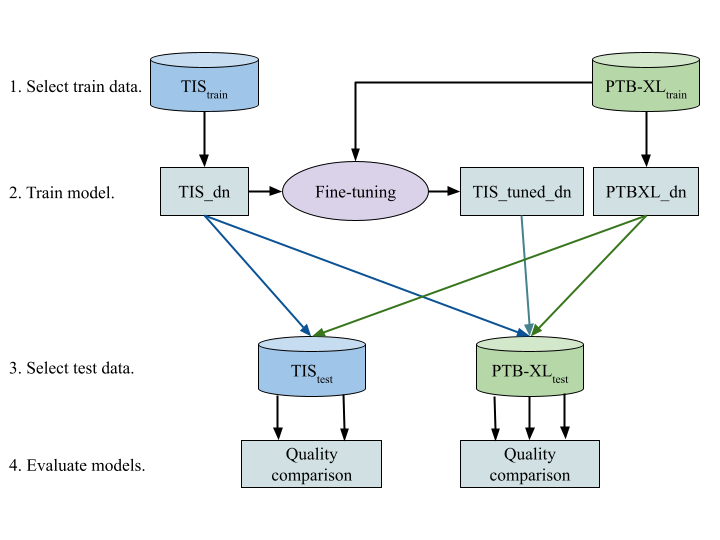}
\caption{Scheme of experiments to compare TIS\_dn, PTBXL\_dn, and TIS\_tuned\_dn.}
\label{fig:Scheme of experiments}
\end{figure}

\begin{table*}[ht]
\caption{Summary of TIS and PTB-XL datasets: Abnormality prevalence \\ for training and validation, and test datasets}
\label{dataset_table}
\centering
\begin{threeparttable}
    \centering
    \begin{tabular}{|c|cc|cc|}
        \hline
        \multirow{3}{*}{\textbf{Pathology}} & \multicolumn{2}{c|}{\textbf{TIS dataset}} & \multicolumn{2}{c|}{\textbf{PTB-XL dataset}} \\ \cline{2-5} 
 & \textbf{\begin{tabular}[c]{@{}c@{}}Train+Val \\ (n=549,279 samples)\end{tabular}} & \textbf{\begin{tabular}[c]{@{}c@{}}Test\\ (n=31,872 samples)\end{tabular}} & \textbf{\begin{tabular}[c]{@{}c@{}}Train+Val\\ (n=17,267 samples)\end{tabular}} & \textbf{\begin{tabular}[c]{@{}c@{}}Test\\ (n=4,347 samples)\end{tabular}} \\ \hline
\textbf{RBBB} & 32,465 (5.9\%) & 1,760 (5.5\%) & 1,307 (7.6\%) & 326 (7.5\%) \\
\textbf{STACH} & 34,838 (6.3\%) & 2,369 (7.4\%) & 656 (3.8\%) & 164 (3.8\%) \\
\textbf{SBRAD} & 25,834 (4.7\%) & 1,419 (4.5\%) & 503 (2.9\%) & 128 (2.9\%)\\
\textbf{AFIB} & 14,942 (2.7\%) & 790 (2.5\%) & 1,198 (6.9\%) & 294 (6.8\%) \\
\textbf{PVC} & 10,090 (1.8\%) & 611 (1.9\%) & 908 (5.3\%) & 226 (5.2\%) \\
\textbf{LBBB} & 4,828 (0.9\%) & 246 (0.8\%) & 483 (2.8\%) & 120 (2.8\%) \\
\textbf{1AVB} & 9,465 (1.7\%) & 488 (1.5\%) & 628 (3.6\%) & 158 (3.6\%) \\
 \hline
\end{tabular}
    \begin{tablenotes}
        \raggedright
        \item[]There are 428,450 (78.0\%) and 12,499 (72.3\%) records in train+val sets of TIS and PTB-XL, 24,835 (77.9\%), and 3175 (73.0\%) records in test sets of TIS and PTB-XL, respectively, which doesn't contain considered abnormalities but may have any other heart disease.
    \end{tablenotes}
\end{threeparttable}
\end{table*}

We want to demonstrate how pre-training the DNN on the variety of datasets combined into one affects the prediction results on the other dataset. We train the models on the TIS dataset and use obtained weights to fine-tune them on the PTB-XL dataset. We selected the same architecture to train the neural network and froze the first blocks of the pre-trained DNN. Then the weights of the last layers are randomly initialized and fine-tuned on PTB-XL. We set multiple experiments to find the best size for the number of blocks of DenseNet to freeze as well as the learning rate. The parameters are chosen from the following options: for the number of frozen blocks {5, 7, 9}, for learning rate {0.001, 0.002, 0.003}. The best metrics on the PTB-XL test dataset are achieved when 7 blocks of DenseNet are frozen and the learning rate is equal to 0.003.

\section{Experiments and results}

We set up experiments to compare the effect of the training datasets on ECG classification. We use both test datasets to evaluate neural networks for each selected abnormality trained exclusively on the observed dataset. Furthermore, we use PTB-XL test data to investigate the effect of pre-training the DNN on the large dataset.

\begin{table*}[t]
\caption{Comparison of the Sens., Spec., G-mean and F2-score metrics of PTBXL\_dn and TIS\_dn DNNs \\ trained to detect one of the 7 abnormalities. DNNs are tested on TIS test data.}
\label{table:DenseNet1}
\resizebox{\textwidth}{!}{
\begin{threeparttable}[!t]
    \centering
    \begin{tabular}{|c|cccccccccccccccc|}
        \hline
        \multicolumn{1}{|c|}{\textbf{Dataset}} & \multicolumn{16}{c|}{\textbf{Pathology}} \\ \hline
\multirow{2}{*}{\textbf{Train}} &  \multicolumn{4}{c|}{\textbf{AFIB}} & \multicolumn{4}{c|}{\textbf{1AVB}} & \multicolumn{4}{c|}{\textbf{RBBB}} & \multicolumn{4}{c|}{\textbf{LBBB}} \\
 &  \textbf{Sens.} & \textbf{Spec.} & \textbf{G-mean} & \multicolumn{1}{c|}{\textbf{F$_{2}$-score}} & \textbf{Sens.} & \textbf{Spec.} & \textbf{G-mean} & \multicolumn{1}{c|}{\textbf{F$_{2}$-score}} & \textbf{Sens.} & \textbf{Spec.} & \textbf{G-mean} & \multicolumn{1}{c|}{\textbf{F$_{2}$-score}} & \textbf{Sens.} & \textbf{Spec.} & \textbf{G-mean} & \multicolumn{1}{c|}{\textbf{F$_{2}$-score}} \\ \hline
PTBXL\_dn &  0.956 & 0.980 & 0.968 & \multicolumn{1}{c|}{0.832} & 0.750 & \textbf{0.940} & 0.840 & \multicolumn{1}{c|}{0.435} & 0.913 & 0.907 & 0.910 & \multicolumn{1}{c|}{0.702} & 0.732 & \textbf{0.985} & 0.849 & 0.546 \\
TIS\_dn &  \textbf{0.982} & \textbf{0.991} & \textbf{0.986} & \multicolumn{1}{c|}{\textbf{0.917}} & \textbf{0.963} & 0.933 & \textbf{0.948} & \multicolumn{1}{c|}{\textbf{0.520}} & \textbf{0.950} & \textbf{0.917} & \textbf{0.933} & \multicolumn{1}{c|}{\textbf{0.746}} & \textbf{0.911} & 0.977 & \textbf{0.943} & \textbf{0.582} \\ \hline
\multicolumn{2}{c}{} &
    \end{tabular} 
\centering
\resizebox{\textwidth}{!}{
    \begin{tabular}{|c|cccccccccccc|}
    \hline
    \multicolumn{1}{|c|}{\textbf{Dataset}} & \multicolumn{12}{c|}{\textbf{Pathology}} \\ \hline
    \multirow{2}{*}{\textbf{Train}} & \multicolumn{4}{c|}{\textbf{STACH}} & \multicolumn{4}{c|}{\textbf{SBRAD}} & \multicolumn{4}{c|}{\textbf{PVC}} \\
     &  \textbf{Sens.} & \textbf{Spec.} & \textbf{G-mean} & \multicolumn{1}{c|}{\textbf{F$_{2}$-score}} & \textbf{Sens.} & \textbf{Spec.} & \textbf{G-mean} & \multicolumn{1}{c|}{\textbf{F$_{2}$-score}} & \textbf{Sens.} & \textbf{Spec.} & \textbf{G-mean} & \multicolumn{1}{c|}{\textbf{F$_{2}$-score}} \\ \hline
    PTBXL\_dn & 0.504 & \textbf{0.980} & 0.703 & \multicolumn{1}{c|}{0.530} & 0.538 & \textbf{0.954} & 0.716 & \multicolumn{1}{c|}{0.487} & 0.931 & 0.931 & 0.931 & 0.551 \\
    TIS\_dn & \textbf{0.977} & 0.937 & \textbf{0.957} & \multicolumn{1}{c|}{\textbf{0.847}} & \textbf{0.987} & 0.945 & \textbf{0.966} & \multicolumn{1}{c|}{\textbf{0.801}} & \textbf{0.987} & \textbf{0.984} & \textbf{0.986} & \textbf{0.852} \\ \hline
    \end{tabular}
}
    \begin{tablenotes}
        \raggedright
        \item[] The best scores are represented by bold values.
    \end{tablenotes}
\end{threeparttable}
}
\end{table*}

We compare three neural networks with the same architectures: the network pre-trained on TIS and fine-tuned on PTB-XL (TIS\_tuned\_dn), and the ones trained either on PTB-XL (PTBXL\_dn) or TIS (TIS\_dn). The scheme of the performed experiments is shown in Fig.~\ref{fig:Scheme of experiments}.

We selected 7 abnormalities according to SCP-ECG standards to train the DNNs: Atrial FIBrillation (AFIB), Right Bundle Branch Block (RBBB), Sinus TACHycardia (STACH), Sinus BRADycardia (SBRAD), first-degree AV Block (1AVB), Left Bundle Branch Block (LBBB), and Premature Ventricular Complexes (PVC). We should note that we merged Complete and Incomplete Left Bundle Branch Blocks (CLBBB and ILBBB) into one disease: the Left Bundle Branch Block (LBBB). Similarly, we merged Complete and Incomplete Right Bundle Branch Blocks into the Right Bundle Branch Block (RBBB). These abnormalities are common and widely represented in each of the considered datasets. The distribution of abnormalities for both datasets is summarized in Table~\ref{dataset_table}.

The prediction results of the networks on TIS and PTB-XL test datasets are presented in Table~\ref{table:DenseNet1} and ~\ref{table:DenseNet2}, respectively. We chose Sensitivity (Sens.), and Specificity (Spec.) metrics to compare the quality of the DNNs. These are widely used metrics for ECG analysis~\cite{berkaya2018survey} Sensitivity assesses how well the model finds people with heart diseases. Specificity gives an estimate of how accurately the model detects people without a disease. We also count G-mean and F2-score to understand which neural network performs better. Higher values of the considered metrics would indicate which model gives more accurate abnormality prediction.

\begin{table*}[ht]
\caption{Comparison of the Sens., Spec., G-mean and F2-score metrics of PTBXL\_dn, TIS\_dn and TIS\_tuned\_dn DNNs trained to detect one of the 7 abnormalities. DNNs are tested on PTB-XL test data.}
\label{table:DenseNet2}
\resizebox{\textwidth}{!}{
\begin{threeparttable}[!t]
    \centering
    \begin{tabular}{|c|cccccccccccccccc|}
        \hline
        \multicolumn{1}{|c|}{\textbf{Dataset}} & \multicolumn{16}{c|}{\textbf{Pathology}} \\ \hline
\multirow{2}{*}{\textbf{Train}} &  \multicolumn{4}{c|}{\textbf{AFIB}} & \multicolumn{4}{c|}{\textbf{1AVB}} & \multicolumn{4}{c|}{\textbf{RBBB}} & \multicolumn{4}{c|}{\textbf{LBBB}} \\
 &  \textbf{Sens.} & \textbf{Spec.} & \textbf{G-mean} & \multicolumn{1}{c|}{\textbf{F$_{2}$-score}} & \textbf{Sens.} & \textbf{Spec.} & \textbf{G-mean} & \multicolumn{1}{c|}{\textbf{F$_{2}$-score}} & \textbf{Sens.} & \textbf{Spec.} & \textbf{G-mean} & \multicolumn{1}{c|}{\textbf{F$_{2}$-score}} & \textbf{Sens.} & \textbf{Spec.} & \textbf{G-mean} & \textbf{F$_{2}$-score} \\ \hline
PTBXL\_dn & 0.929 & 0.969 & 0.949 & \multicolumn{1}{c|}{0.868} & 0.905 & 0.884 & 0.894 & \multicolumn{1}{c|}{0.566} & 0.972 & 0.942 & 0.957 & \multicolumn{1}{c|}{0.855} & 0.842 & 0.980 & 0.908 & 0.762 \\
TIS\_dn & \textbf{0.952} & 0.953 & 0.953 & \multicolumn{1}{c|}{0.850} & \textbf{0.918} & 0.859 & 0.888 & \multicolumn{1}{c|}{0.531} & \textbf{0.985} & 0.938 & 0.961 & \multicolumn{1}{c|}{0.856} & \textbf{0.992} & 0.907 & 0.948 & 0.599 \\
TIS\_tuned\_dn & 0.942 & \textbf{0.972} & \textbf{0.957} & \multicolumn{1}{c|}{\textbf{0.884}} & 0.892 & \textbf{0.923} & \textbf{0.908} & \multicolumn{1}{c|}{\textbf{0.644}} & \textbf{0.985} & \textbf{0.947} & \textbf{0.966} & \multicolumn{1}{c|}{\textbf{0.873}} & 0.917 & \textbf{0.982} & \textbf{0.949} & \textbf{0.827} \\ \hline
\multicolumn{2}{c}{} &
    \end{tabular} 
\centering
\resizebox{\textwidth}{!}{
\begin{tabular}{|c|cccccccccccc|}
\hline
\multicolumn{1}{|c|}{\textbf{Dataset}} & \multicolumn{12}{c|}{\textbf{Pathology}} \\ \hline
\multirow{2}{*}{\textbf{Train}} &  \multicolumn{4}{c|}{\textbf{STACH}} & \multicolumn{4}{c|}{\textbf{SBRAD}} & \multicolumn{4}{c|}{\textbf{PVC}} \\
 & \textbf{Sens.} & \textbf{Spec.} & \textbf{G-mean} & \multicolumn{1}{c|}{\textbf{F$_{2}$-score}} & \textbf{Sens.} & \textbf{Spec.} & \textbf{G-mean} & \multicolumn{1}{c|}{\textbf{F$_{2}$-score}} & \textbf{Sens.} & \textbf{Spec.} & \textbf{G-mean} & \textbf{F$_{2}$-score} \\ \hline
PTBXL\_dn & 0.951 & 0.962 & 0.957 & \multicolumn{1}{c|}{0.804} & 0.789 & \textbf{0.928} & 0.856 & \multicolumn{1}{c|}{0.551} & 0.956 & 0.953 & 0.954 & 0.821 \\
TIS\_dn & \textbf{0.982} & 0.906 & 0.943 & \multicolumn{1}{c|}{0.665} & \textbf{0.961} & 0.848 & \textbf{0.902} & \multicolumn{1}{c|}{0.481} & \textbf{0.991} & 0.970 & 0.981 & 0.895 \\
TIS\_tuned\_dnL & 0.963 & \textbf{0.979} & \textbf{0.971} & \multicolumn{1}{c|}{\textbf{0.876}} & 0.891 & 0.913 & \textbf{0.902} & \multicolumn{1}{c|}{\textbf{0.575}} & 0.987 & \textbf{0.980} & \textbf{0.983} & \textbf{0.921} \\ \hline
\end{tabular}
}
    \begin{tablenotes}
        \raggedright
        \item[] The best scores are represented by bold values.
    \end{tablenotes}
\end{threeparttable}

}
\end{table*}

Comparison of the DNNs demonstrates the importance of large and diverse train data. TIS\_dn models generalize disease classification significantly better than PTBXL\_dn models and are stable regardless of the test dataset. For most of the observed abnormalities, TIS\_dn models show comparable quality on the PTB-XL test dataset to PTBXL\_dn models. On the contrary, the metrics of the PTBXL\_dn decrease on the TIS test dataset. Different training data have a smaller impact on prediction quality for abnormalities that have less variety among patients. Therefore, the metrics of the PTBXL\_dn DNNs detecting atrial fibrillation or right bundle branch block demonstrate a moderate change. However, the difference becomes significant for other heart diseases such as 1AVB, PVC, or LBBB and reaches up to 10\% for G-mean.

TIS\_tuned\_dn networks give promising results. Considering the observed metrics, a TIS\_tuned\_dn gives better predictions as compared with both TIS\_dn or PTBXL\_dn networks for all the abnormalities. The G-mean metric is increased by up to 2\% compared to TIS\_dn and up to 4\% compared to PTBXL\_dn, which is a significant improvement in the quality of ECG classification. Moreover, if G-mean does not have a significant improvement, F2-score increases dramatically, which means that a TIS\_tuned\_dn can much better detect people with abnormalities.

\begin{table*}[ht]
\caption{Comparison of PTBXL\_dn, TIS\_dn, hTIS\_dn and hTIS\_tuned\_dn DNNs to detect atrial \\ fibrillation (AFIB) and premature ventricular complexes (PVC). The performance \\ of the DNNs is presented for the PTB-XL test data.}
\label{table:ReducedResults}
\centering
\begin{threeparttable}
    \begin{tabular}{|cc|cccc|cccc|}
        \hline
        \multicolumn{2}{|c|}{\textbf{Dataset}}& \multicolumn{8}{c|}{\textbf{Pathology}}\\ \hline
\multirow{2}{*}{\textbf{Train}} & \multirow{2}{*}{\textbf{Number of Train samples}} & \multicolumn{4}{c|}{\textbf{AFIB}} & \multicolumn{4}{c|}{\textbf{PVC}} \\
& & \textbf{Sens.} & \textbf{Spec.} & \textbf{G-mean} & \textbf{F$_{2}$-score} & \textbf{Sens.} & \textbf{Spec.} & \textbf{G-mean} & \textbf{F$_{2}$-score} \\ \hline
PTBXL\_dn  & 17,267 & 0.929 & 0.969  & 0.949 & 0.868 & 0.956  & 
0.953 & 0.954 & 0.821 \\
TIS\_dn &  549,278 & \textbf{0.953}& 0.952 & 0.953 & 0.850 & 0.991 & \textbf{0.970} & 0.981 & 0.895 \\
hTIS\_dn  & 274,639 & 0.946 & 0.967 & \textbf{0.956}  &
 0.876 & 0.987 & 0.934 & 0.960 & 0.796 \\
hTIS\_tuned\_dn  & 291,906 & 0.935 & \textbf{0.976} & \textbf{0.956} & \textbf{0.888} & \textbf{0.996} & \textbf{0.970} & \textbf{0.983} & \textbf{0.897} \\ \hline
    \end{tabular} 
    \begin{tablenotes}
        \raggedright
        \item[] The best scores are represented by bold values.
    \end{tablenotes}
\end{threeparttable}

\end{table*}

We also trained several neural networks with different architectures apart from DenseNet. We conducted the same experiments with various neural networks used as benchmarks for the classification of the PTB-XL dataset ~\cite{strodthoff2020deep} and achieved the same results. The results of the experiments for these DNNs are presented in Appendix Tables~\ref{table:ResNet18_1} -~\ref{table:LSTM_2}.

The TIS dataset is large, which affects the quality of classification and helps to generalize the DNNs. However, we demonstrate that there are other characteristics of each dataset apart from the size that influence the DNN prediction performance. We run the experiments halving the size of the TIS training dataset with the maintaining proportions of abnormality prevalence to show that fine-tuning on relatively small PTB-XL will still improve the classification of PTB-XL test data. We selected 2 abnormalities AFIB and PVC and trained the DenseNet model on a halved TIS train dataset (hTIS\_dn). Then we froze the layers of pre-trained networks and fine-tuned them on PTB-XL (hTIS\_tuned\_dn), which is almost 30 times smaller than TIS.

The comparison of the DNNs trained on datasets of different sizes on PTB-XL test data is presented in Table~\ref{table:ReducedResults}. The metrics are not improved for the hTIS\_dn model compared to PTBXL\_dn and TIS\_dn. Wherein, the metrics of hTIS\_tuned\_dn models are the best among the observed neural networks. Despite halving the training dataset and reducing the number of samples by almost 300,000, fine-tuning the model with approximately 17,000 samples from another data source improves the metrics on PTB-XL. We demonstrate that a model trained on a much smaller but specific dataset showed higher quality on the selected test data compared to the models trained on the full-size dataset.

\begin{table*}[!htb]
\caption{Prediction performance of PTBXL\_dn and TIS\_tuned\_dn DNNs compared with \\ the performance of the doctors from different hospitals.}
\label{table:Doctors}
\centering
\begin{tabular}{|c|cccc|cccc|}
\hline
\multirow{2}{*}{\textbf{Prediction}} & \multicolumn{4}{c|}{\textbf{AFIB}} & \multicolumn{4}{c|}{\textbf{PVC}} \\
 & \textbf{Sens.} & \textbf{Spec.} & \textbf{G-mean} & \textbf{F$_{2}$-score} & \textbf{Sens.} & \textbf{Spec.} & \textbf{G-mean} & \textbf{F$_{2}$-score} \\ \hline
PTBXL\_dn & 0.941 & 0.961 & 0.951 & 0.860 & 0.962 & 0.956 & 0.959 & 0.833 \\
TIS\_tuned\_dn & 0.941 & 0.966 & 0.953 & 0.870 & 1.000 & 0.981 & 0.990 & 0.935 \\ \hline
Doctor 1 & 0.941 & 0.974 & 0.958 & 0.889 & 0.923 & 0.979 & 0.951 & 0.870 \\
Doctor 2 & 0.971 & 0.899 & 0.934 & 0.764 & 0.923 & 0.992 & 0.957 & 0.909 \\
Doctor 3 & 0.912 & 0.981 & 0.946 & 0.881 & 0.962 & 0.958 & 0.960 & 0.839 \\ \hline
\end{tabular}
\end{table*}

To estimate the quality of DNNs in clinical practice, we compared their prediction with the independent ECG evaluation by three doctors from different hospitals with years of experience in reading and diagnosing ECG records. We took a random subset of 500 samples from the PTB-XL test dataset with the maintaining proportions of the classes. Each cardiologist annotated these 500 records selecting whether two abnormalities observed in this paper (AFIB and PVC) are present in the ECG record. The subset of PTB-XL had 34 and 26 ECG records with AFIB and PVC respectively. After the evaluation, we compared the annotation of the doctors with the PTBXL\_dn and TIS\_tuned\_dn DNNs.

The prediction performance of Densenet networks and the doctors' annotation on a subset of 500 ECGs are presented in Table~\ref{table:Doctors}.  We also plot sensitivity/specificity curves in Fig.~\ref{fig:doctor_metrics} and confusion matrices of the doctor's annotation and the DNNs prediction in Appendix Fig.~\ref{fig:cm for doctors and DNNs PVC},~\ref{fig:cm for doctors and DNNs AFIB}. The metrics demonstrate that TIS\_tuned\_dn outperforms 2 cardiologists for AFIB and all the cardiologists for PVC.

\begin{figure}[!t]{}
     \centering
     \begin{subfigure}{\columnwidth}
         \centering
         \includegraphics[width=\columnwidth]{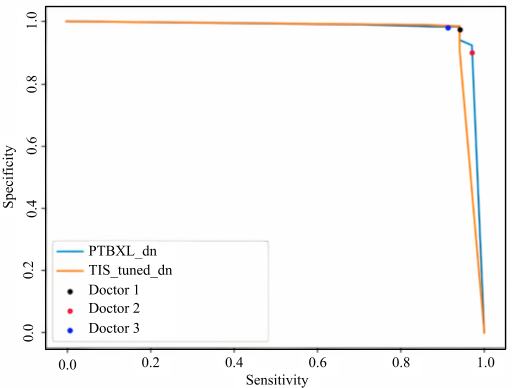}
         \subcaption{AFIB}
         \label{fig:AFIB}
     \end{subfigure}
     \hfill
     \begin{subfigure}{\columnwidth}
         \centering
         \includegraphics[width=\columnwidth]{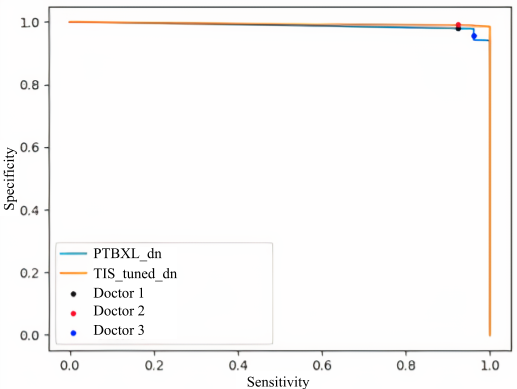}
         \subcaption{PVC}
         \label{fig:PVC}
     \end{subfigure}
     \caption{Sensitivity-specificity curves show the prediction quality for PTBXL\_dn and TIS\_tuned\_dn networks to detect atrial fibrillation (AFIB) and premature ventricular complexes (PVC). Points correspond to the performance of three different doctors' annotations.}
     \label{fig:doctor_metrics}
\end{figure}

\section{Discussion}

Despite many convincing studies on 12-lead ECG classification, the research rarely addresses the problem of generalization of the neural networks. This is primarily since usually there is no opportunity to train the DNNs on large amounts of data from various sources. Most studies use the same train and test dataset for the ECG analysis~\cite{berkaya2018survey}, and there are a few researchers that use different datasets~\cite{zhang2021mlbf,li2021automatic}. However, they are trained and validated on the same dataset. This eliminates the possibility of correctly validating the DNNs on ECGs from various sources because records may differ for multiple reasons. For instance, data can be initially preprocessed during the ECG recording or later during the doctors’ annotation. Another possible reason that can cause the difference in the quality of records is the diversity of ECG devices that can record ECGs with specific parameters, filtering, or distinct noise.

Large and heterogeneous ECG data help with networks' generalization. We demonstrate that DNNs trained on the collection of different datasets maintain high metrics for different datasets for the observed heart diseases and show metrics comparable to clinical quality on data from another domain for two selected abnormalities. DNNs trained on open datasets can achieve good initial results for several cardiac abnormalities, but they are unstable for data from different domains. It is necessary to use training data from different sources to be able to generalize the neural network. Another attempt to make the network more stable is to create a unified approach to filtering and preprocessing during data collection. Most of the open data is initially preprocessed, which may affect the loss of features that determine the abnormality of the record. One of the ways to solve this problem can be publishing raw ECG signals in addition to preprocessed data.

An important observation from the experiments in this paper is that neural networks trained on many various data can improve in quality with a relatively small increase in the training data. These results allow us to fine-tune the DNNs on datasets from different sources to get better prediction performance. Thus, researchers who do not have large amounts of data can achieve high-quality results on relatively small amounts of data.

The good prediction performance of the DNNs trained to diagnose 12-lead ECGs does not limit the potential for improvement. As the next steps to develop existing classification methods, new parameters, which can lead to the better generalization of the neural networks, must be considered. First of all, these are patient metadata, which will give additional information for certain records. There are already some developments in analyzing metadata~\cite{kwon2020deep}, however, the analysis has limitations. The reason for that is a small amount of additional information about a record due to security reasons or the lack of a single system for storing patients' data. One of the solutions to the problem associated with the confidentiality of storing patients' metadata can be an identification of a person by the ECG~\cite{zhang2021human}.

To improve the abnormality prediction models, it will be helpful to develop algorithms for each disease individually. In some cases, there is no need to use the entire ECG record when only a few parts are sufficient. This approach can simplify the neural networks by not providing the parts of the record that are not affected by the considered abnormality. In addition, algorithms should be tested on a variety of real-time data from different sources with feedback from cardiologists. This will help in analyzing incorrectly predicted samples and early identification of the errors in the algorithm.

The ability to fine-tune the DNNs pre-trained on a large dataset allows us to work on the described improvements. Researchers can use the networks as backbones for more complex architectures to achieve the best quality. ECG metadata and hand-crafted features can be used with the DNNs to evaluate their influence on the prediction. The weights of the neural networks can also be used for transfer learning for studies that do not have a lot of labeled data, for instance, analysis of ECG records to detect hyperkalemia, hypokalemia~\cite{lin2020deep}, or human emotions~\cite{sarkar2020self}. Moreover, trained networks can be used as benchmarks for datasets from different sources and provide the possibility to verify new models.

\section{Conclusion}

Deep neural networks have achieved promising results in predicting heart diseases with ECG records. However, due to the difference in ECG records from multiple data sources and the small amount of available data for training, current DNNs are not generalized to be widely applied in practice.

In this paper, we present the methodology to achieve better abnormality prediction quality regardless of the dataset. We demonstrate that training neural networks on a variety of datasets with further fine-tuning on the specific dataset shows better quality than DNN trained on this specific data set only. This approach can help the application of DNNs for ECGs analysis in medical centers that have different data sources. In addition, it can reduce both the cost and time spent on data annotation by cardiologists, since this approach requires less labeled data.

We also show the significance of a large diverse data for a generalization of the neural networks. The improvement of generalization ability leads to the more efficient use of neural networks in real-life applications. We demonstrate that current DNNs trained on a large collection of different datasets and fine-tuned for the selected dataset can achieve cardiologist-level performance. The proposed methodology can be applied to new algorithms and architectures, thereby providing better prediction quality.

For further improvement, we consider the analysis of ECG metadata for abnormality classification. Considering additional information about the patient can be substantial and give more accurate predictions. Moreover, we consider only patients that are older than 18 years. This limitation has to be considered when implementing new methods for ECG classification due to criteria for diagnosing children’s and adults’ heart diseases may differ, which can cause the inability of trained DNNs to predict the abnormalities accurately. One of the approaches would be training different models for predicting cardiac abnormalities in children and adults.

\appendix
\label{appendix:1}

\begin{figure}[t]
     \centering
     \begin{subfigure}{0.17\textwidth}
         \centering
         \includegraphics[width=\textwidth]{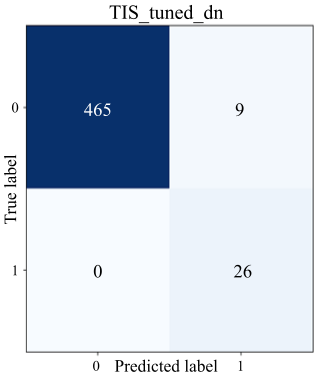}
    \end{subfigure} 
    \begin{subfigure}{0.17\textwidth}
         \centering
         \includegraphics[width=\textwidth]{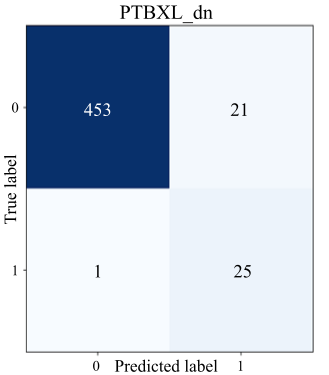}
    \end{subfigure} 
    \par\bigskip
    
     \centering
     \begin{subfigure}{0.17\textwidth}
         \centering
         \includegraphics[width=\textwidth]{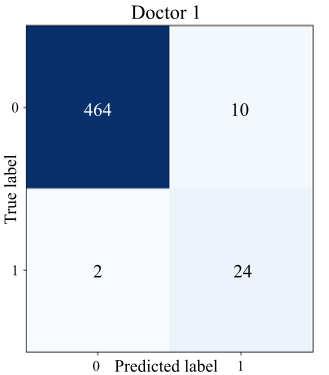}
    \end{subfigure}
    \begin{subfigure}{0.17\textwidth}
         \centering
         \includegraphics[width=\textwidth]{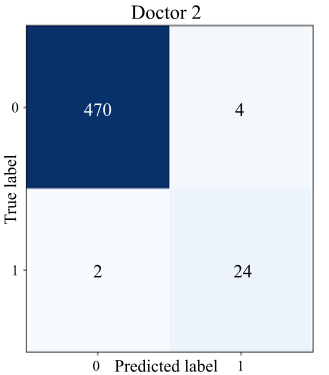}
    \end{subfigure}
    \par\bigskip
    \begin{subfigure}{0.17\textwidth}
         \centering
         \includegraphics[width=\textwidth]{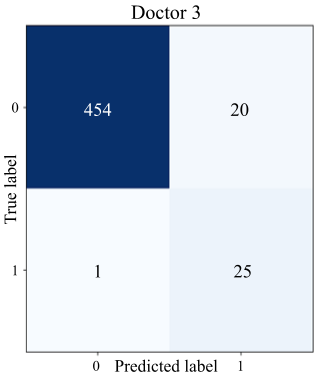}
    \end{subfigure}
    \caption{Confusion matrices for doctors and DNNs for PVC.}
\label{fig:cm for doctors and DNNs PVC}
\end{figure}

\begin{figure}[hb]
     \centering
     \begin{subfigure}{0.17\textwidth}
         \centering
         \includegraphics[width=\textwidth]{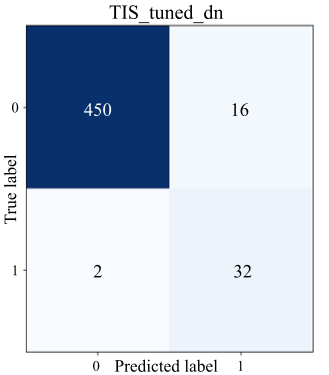}
    \end{subfigure} 
    \begin{subfigure}{0.17\textwidth}
         \centering
         \includegraphics[width=\textwidth]{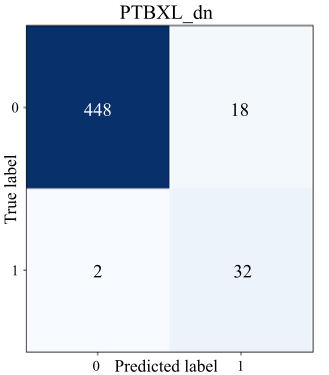}
    \end{subfigure} 
    \par\bigskip
    
     \centering
     \begin{subfigure}{0.17\textwidth}
         \centering
         \includegraphics[width=\textwidth]{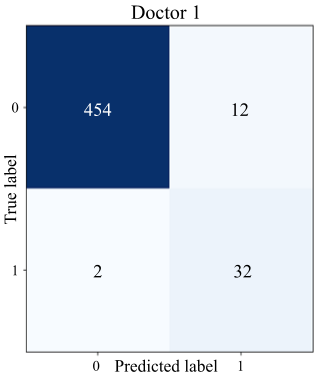}
    \end{subfigure}
    \begin{subfigure}{0.17\textwidth}
         \centering
         \includegraphics[width=\textwidth]{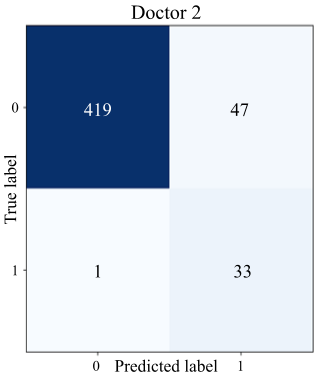}
    \end{subfigure}
    \par\bigskip
    \begin{subfigure}{0.17\textwidth}
         \centering
         \includegraphics[width=\textwidth]{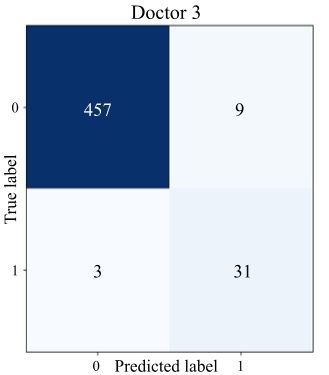}
    \end{subfigure}
    \caption{Confusion matrices for doctors and DNNs for AFIB.}
\label{fig:cm for doctors and DNNs AFIB}
\end{figure}

\onecolumn
\begin{table*}[t]
\caption{Comparison of the metrics of ResNet1d18 networks: trained either on PTB-XL (PTBXL\_rn18) or TIS (TIS\_rn18) to detect one of the 7 abnormalities. DNNs are tested on TIS test data.}
\label{table:ResNet18_1}
\resizebox{\textwidth}{!}{
\begin{threeparttable}
    \centering
    \begin{tabular}{|c|cccccccccccccccc|}
        \hline
        \multicolumn{1}{|c|}{\textbf{Dataset}} & \multicolumn{16}{c|}{\textbf{Pathology}} \\ \hline
\multirow{2}{*}{\textbf{Train}} &  \multicolumn{4}{c|}{\textbf{AFIB}} & \multicolumn{4}{c|}{\textbf{1AVB}} & \multicolumn{4}{c|}{\textbf{RBBB}} & \multicolumn{4}{c|}{\textbf{LBBB}} \\
 &  \textbf{Sens.} & \textbf{Spec.} & \textbf{G-mean} & \multicolumn{1}{c|}{\textbf{F$_{2}$-score}} & \textbf{Sens.} & \textbf{Spec.} & \textbf{G-mean} & \multicolumn{1}{c|}{\textbf{F$_{2}$-score}} & \textbf{Sens.} & \textbf{Spec.} & \textbf{G-mean} & \multicolumn{1}{c|}{\textbf{F$_{2}$-score}} & \textbf{Sens.} & \textbf{Spec.} & \textbf{G-mean} & \multicolumn{1}{c|}{\textbf{F$_{2}$-score}} \\ \hline
PTBXL\_rn18 & 0.972 & 0.981 & 0.977 & \multicolumn{1}{c|}{0.851} & 0.686 & \textbf{0.963} & 0.813 & \multicolumn{1}{c|}{0.487} & 0.792 & \textbf{0.960} & 0.872 & \multicolumn{1}{c|}{0.723} & 0.780 & \textbf{0.990} & 0.879 & \multicolumn{1}{c|}{\textbf{0.642}} \\
TIS\_rn18 & \textbf{0.976} & \textbf{0.993} & \textbf{0.984} & \multicolumn{1}{c|}{\textbf{0.929}} & \textbf{0.957} & 0.931 & \textbf{0.944} & \multicolumn{1}{c|}{\textbf{0.511}} & \textbf{0.967} & 0.902 & \textbf{0.934} & \multicolumn{1}{c|}{\textbf{0.727}} & \textbf{0.919} & 0.959 & \textbf{0.939} & \multicolumn{1}{c|}{0.450} \\ \hline
\multicolumn{2}{c}{} &
    \end{tabular} 
\centering
\resizebox{\textwidth}{!}{
\begin{tabular}{|c|cccccccccccc|}
\hline
\multicolumn{1}{|c|}{\textbf{Dataset}} & \multicolumn{12}{c|}{\textbf{Pathology}} \\ \hline
\multirow{2}{*}{\textbf{Train}} & \multicolumn{4}{c|}{\textbf{STACH}} & \multicolumn{4}{c|}{\textbf{SBRAD}} & \multicolumn{4}{c|}{\textbf{PVC}} \\
 &  \textbf{Sens.} & \textbf{Spec.} & \textbf{G-mean} & \multicolumn{1}{c|}{\textbf{F$_{2}$-score}} & \textbf{Sens.} & \textbf{Spec.} & \textbf{G-mean} & \multicolumn{1}{c|}{\textbf{F$_{2}$-score}} & \textbf{Sens.} & \textbf{Spec.} & \textbf{G-mean} & \multicolumn{1}{c|}{\textbf{F$_{2}$-score}} \\ \hline
PTBXL\_rn18 & 0.488 & \textbf{0.984} & 0.693 & \multicolumn{1}{c|}{0.520} & 0.545 & \textbf{0.947} & 0.718 & \multicolumn{1}{c|}{0.480} & 0.923 & 0.978 & 0.950 & \multicolumn{1}{c|}{0.762} \\
TIS\_rn18 & \textbf{0.981} & 0.931 & \textbf{0.956} & \multicolumn{1}{c|}{\textbf{0.841}} & \textbf{0.989} & 0.944 & \textbf{0.966} & \multicolumn{1}{c|}{\textbf{0.798}} & \textbf{0.993} & \textbf{0.983} & \textbf{0.988} & \multicolumn{1}{c|}{\textbf{0.845}} \\ \hline
\end{tabular}
}
    \begin{tablenotes}
        \raggedright
        \item[] The best scores are represented by bold values.
    \end{tablenotes}
\end{threeparttable}

}
\end{table*}

\begin{table*}[t]
\caption{Comparison of the metrics of three ResNet1d18 networks: pre-trained on TIS and fine-tuned on PTB-XL (TIS\_tuned\_rn18), and trained either on PTB-XL (PTBXL\_rn18) or TIS (TIS\_rn18) to detect one of the 7 abnormalities. DNNs are tested on PTB-XL test data.}
\label{table:ResNet18_2}
\resizebox{\textwidth}{!}{
\begin{threeparttable}
    \centering
    \begin{tabular}{|c|cccccccccccccccc|}
        \hline
        \multicolumn{1}{|c|}{\textbf{Dataset}} & \multicolumn{16}{c|}{\textbf{Pathology}} \\ \hline
\multirow{2}{*}{\textbf{Train}} &  \multicolumn{4}{c|}{\textbf{AFIB}} & \multicolumn{4}{c|}{\textbf{1AVB}} & \multicolumn{4}{c|}{\textbf{RBBB}} & \multicolumn{4}{c|}{\textbf{LBBB}} \\
 &  \textbf{Sens.} & \textbf{Spec.} & \textbf{G-mean} & \multicolumn{1}{c|}{\textbf{F$_{2}$-score}} & \textbf{Sens.} & \textbf{Spec.} & \textbf{G-mean} & \multicolumn{1}{c|}{\textbf{F$_{2}$-score}} & \textbf{Sens.} & \textbf{Spec.} & \textbf{G-mean} & \multicolumn{1}{c|}{\textbf{F$_{2}$-score}} & \textbf{Sens.} & \textbf{Spec.} & \textbf{G-mean} & \textbf{F$_{2}$-score} \\ \hline
PTBXL\_rn18 & \textbf{0.946} & 0.965 & \textbf{0.955} & \multicolumn{1}{c|}{0.871} & 0.880 & \textbf{0.925} & 0.902 & \multicolumn{1}{c|}{\textbf{0.639}} & 0.914 & \textbf{0.973} & 0.943 & \multicolumn{1}{c|}{\textbf{0.870}} & 0.908 & \textbf{0.983} & 0.945 & \multicolumn{1}{c|}{0.825} \\
TIS\_rn18 & \textbf{0.946} & 0.963 & 0.954 & \multicolumn{1}{c|}{0.866} & 0.943 & 0.847 & 0.894 & \multicolumn{1}{c|}{0.524} & \textbf{0.994} & 0.931 & 0.962 & \multicolumn{1}{c|}{0.850} & \textbf{0.992} & 0.873 & 0.931 & \multicolumn{1}{c|}{0.524} \\
TIS\_tuned\_rn18 & 0.939 & \textbf{0.972} & \textbf{0.955} & \multicolumn{1}{c|}{\textbf{0.881}} & \textbf{0.962} & 0.887 & \textbf{0.924} & \multicolumn{1}{c|}{0.605} & 0.985 & 0.942 & \textbf{0.963} & \multicolumn{1}{c|}{0.864} & 0.925 & 0.982 & \textbf{0.953} & \multicolumn{1}{c|}{\textbf{0.830}} \\ \hline
\multicolumn{2}{c}{} &
    \end{tabular} 
\centering
\resizebox{\textwidth}{!}{
\begin{tabular}{|c|cccccccccccc|}
\hline
\multicolumn{1}{|c|}{\textbf{Dataset}} & \multicolumn{12}{c|}{\textbf{Pathology}} \\ \hline
\multirow{2}{*}{\textbf{Train}} &  \multicolumn{4}{c|}{\textbf{STACH}} & \multicolumn{4}{c|}{\textbf{SBRAD}} & \multicolumn{4}{c|}{\textbf{PVC}} \\
 & \textbf{Sens.} & \textbf{Spec.} & \textbf{G-mean} & \multicolumn{1}{c|}{\textbf{F$_{2}$-score}} & \textbf{Sens.} & \textbf{Spec.} & \textbf{G-mean} & \multicolumn{1}{c|}{\textbf{F$_{2}$-score}} & \textbf{Sens.} & \textbf{Spec.} & \textbf{G-mean} & \textbf{F$_{2}$-score} \\ \hline
PTBXL\_rn18 & 0.957 & \textbf{0.972} & 0.965 & \multicolumn{1}{c|}{\textbf{0.845}} & 0.844 & 0.900 & 0.872 & \multicolumn{1}{c|}{0.519} & 0.951 & 0.978 & 0.965 & \multicolumn{1}{c|}{0.889} \\
TIS\_rn18 & \textbf{0.994} & 0.897 & 0.944 & \multicolumn{1}{c|}{0.653} & \textbf{0.961} & 0.850 & \textbf{0.904} & \multicolumn{1}{c|}{0.485} & \textbf{0.996} & 0.965 & \textbf{0.980} & \multicolumn{1}{c|}{0.882} \\
TIS\_tuned\_rn18 & 0.976 & 0.956 & \textbf{0.966} & \multicolumn{1}{c|}{0.802} & 0.844 & \textbf{0.936} & 0.889 & \multicolumn{1}{c|}{\textbf{0.607}} & 0.973 & \textbf{0.983} & 0.978 & \multicolumn{1}{c|}{\textbf{0.921}} \\ \hline
\end{tabular}
}
    \begin{tablenotes}
        \raggedright
        \item[] The best scores are represented by bold values.
    \end{tablenotes}
\end{threeparttable}

}
\end{table*}

\begin{table*}[b]
\caption{Comparison of the metrics of ResNet1d50 networks: trained either on PTB-XL (PTBXL\_rn50) or TIS (TIS\_rn50) to detect one of the 7 abnormalities. DNNs are tested on TIS test data.}
\label{table:ResNet50_1}
\resizebox{\textwidth}{!}{
\begin{threeparttable}
    \centering
    \begin{tabular}{|c|cccccccccccccccc|}
        \hline
        \multicolumn{1}{|c|}{\textbf{Dataset}} & \multicolumn{16}{c|}{\textbf{Pathology}} \\ \hline
\multirow{2}{*}{\textbf{Train}} &  \multicolumn{4}{c|}{\textbf{AFIB}} & \multicolumn{4}{c|}{\textbf{1AVB}} & \multicolumn{4}{c|}{\textbf{RBBB}} & \multicolumn{4}{c|}{\textbf{LBBB}} \\
 &  \textbf{Sens.} & \textbf{Spec.} & \textbf{G-mean} & \multicolumn{1}{c|}{\textbf{F$_{2}$-score}} & \textbf{Sens.} & \textbf{Spec.} & \textbf{G-mean} & \multicolumn{1}{c|}{\textbf{F$_{2}$-score}} & \textbf{Sens.} & \textbf{Spec.} & \textbf{G-mean} & \multicolumn{1}{c|}{\textbf{F$_{2}$-score}} & \textbf{Sens.} & \textbf{Spec.} & \textbf{G-mean} & \multicolumn{1}{c|}{\textbf{F$_{2}$-score}} \\ \hline
PTBXL\_rn50 &  0.963 & \textbf{0.997} & 0.980 & \multicolumn{1}{c|}{\textbf{0.947}} & 0.676 & \textbf{0.960} & 0.806 & \multicolumn{1}{c|}{0.468} & 0.874 & \textbf{0.927} & 0.900 & \multicolumn{1}{c|}{0.714} & 0.825 & \textbf{0.984} & 0.901 & \multicolumn{1}{c|}{\textbf{0.602}} \\
TIS\_rn50 &  \textbf{0.973} & 0.992 & \textbf{0.983} & \multicolumn{1}{c|}{0.921} & \textbf{0.977} & 0.920 & \textbf{0.948} & \multicolumn{1}{c|}{\textbf{0.482}} & \textbf{0.956} & 0.915 & \textbf{0.935} & \multicolumn{1}{c|}{\textbf{0.745}} & \textbf{0.911} & 0.971 & \textbf{0.940} & \multicolumn{1}{c|}{0.530} \\ \hline
\multicolumn{2}{c}{} &
    \end{tabular} 
\centering
\resizebox{\textwidth}{!}{
\begin{tabular}{|c|cccccccccccc|}
\hline
\multicolumn{1}{|c|}{\textbf{Dataset}} & \multicolumn{12}{c|}{\textbf{Pathology}} \\ \hline
\multirow{2}{*}{\textbf{Train}} & \multicolumn{4}{c|}{\textbf{STACH}} & \multicolumn{4}{c|}{\textbf{SBRAD}} & \multicolumn{4}{c|}{\textbf{PVC}} \\
 &  \textbf{Sens.} & \textbf{Spec.} & \textbf{G-mean} & \multicolumn{1}{c|}{\textbf{F$_{2}$-score}} & \textbf{Sens.} & \textbf{Spec.} & \textbf{G-mean} & \multicolumn{1}{c|}{\textbf{F$_{2}$-score}} & \textbf{Sens.} & \textbf{Spec.} & \textbf{G-mean} & \multicolumn{1}{c|}{\textbf{F$_{2}$-score}} \\ \hline
PTBXL\_rn50 & 0.496 & \textbf{0.984} & 0.699 & \multicolumn{1}{c|}{0.528} & 0.706 & \textbf{0.948} & 0.818 & \multicolumn{1}{c|}{0.606} & 0.928 & 0.966 & 0.947 & \multicolumn{1}{c|}{0.695} \\
TIS\_rn50 & \textbf{0.979} & 0.939 & \textbf{0.959} & \multicolumn{1}{c|}{\textbf{0.853}} & \textbf{0.984} & 0.946 & \textbf{0.965} & \multicolumn{1}{c|}{\textbf{0.802}} & \textbf{0.992} & \textbf{0.984} & \textbf{0.988} & \multicolumn{1}{c|}{\textbf{0.856}} \\ \hline
\end{tabular}
}
    \begin{tablenotes}
        \raggedright
        \item[] The best scores are represented by bold values.
    \end{tablenotes}
\end{threeparttable}

}
\end{table*}

\begin{table*}[htbp]
\caption{Comparison of the metrics of three ResNet1d50 networks: pre-trained on TIS and fine-tuned on PTB-XL (TIS\_tuned\_rn50), and trained either on PTB-XL (PTBXL\_rn50) or TIS (TIS\_rn50) to detect one of the 7 abnormalities. DNNs are tested on PTB-XL test data.}
\label{table:ResNet50_2}
\resizebox{\textwidth}{!}{
\begin{threeparttable}
    \centering
    \begin{tabular}{|c|cccccccccccccccc|}
        \hline
        \multicolumn{1}{|c|}{\textbf{Dataset}} & \multicolumn{16}{c|}{\textbf{Pathology}} \\ \hline
\multirow{2}{*}{\textbf{Train}} &  \multicolumn{4}{c|}{\textbf{AFIB}} & \multicolumn{4}{c|}{\textbf{1AVB}} & \multicolumn{4}{c|}{\textbf{RBBB}} & \multicolumn{4}{c|}{\textbf{LBBB}} \\
 &  \textbf{Sens.} & \textbf{Spec.} & \textbf{G-mean} & \multicolumn{1}{c|}{\textbf{F$_{2}$-score}} & \textbf{Sens.} & \textbf{Spec.} & \textbf{G-mean} & \multicolumn{1}{c|}{\textbf{F$_{2}$-score}} & \textbf{Sens.} & \textbf{Spec.} & \textbf{G-mean} & \multicolumn{1}{c|}{\textbf{F$_{2}$-score}} & \textbf{Sens.} & \textbf{Spec.} & \textbf{G-mean} & \textbf{F$_{2}$-score} \\ \hline
PTBXL\_rn50 & 0.915 & \textbf{0.987} & 0.950 & \multicolumn{1}{c|}{0.898} & 0.867 & \textbf{0.916} & 0.891 & \multicolumn{1}{c|}{\textbf{0.611}} & 0.951 & 0.952 & 0.951 & \multicolumn{1}{c|}{0.857} & 0.925 & \textbf{0.969} & 0.947 & \multicolumn{1}{c|}{\textbf{0.767}} \\
TIS\_rn50 &  \textbf{0.949} & 0.959 & \textbf{0.954} & \multicolumn{1}{c|}{0.861} & \textbf{0.962} & 0.819 & 0.888 & \multicolumn{1}{c|}{0.493} & \textbf{0.988} & 0.934 & \textbf{0.960} & \multicolumn{1}{c|}{0.851} & \textbf{0.992} & 0.891 & 0.940 & \multicolumn{1}{c|}{0.562}\\
TIS\_tuned\_rn50 & 0.918 & \textbf{0.987} & 0.952 & \multicolumn{1}{c|}{\textbf{0.901}} & 0.930 & 0.893 & \textbf{0.912} & \multicolumn{1}{c|}{0.600} & 0.963 &\textbf{ 0.956} & \textbf{0.960} & \multicolumn{1}{c|}{\textbf{0.875}} & 0.950 & 0.965 & \textbf{0.957} & \multicolumn{1}{c|}{0.766} \\ \hline
\multicolumn{2}{c}{} &
    \end{tabular} 
\centering
\resizebox{\textwidth}{!}{
\begin{tabular}{|c|cccccccccccc|}
\hline
\multicolumn{1}{|c|}{\textbf{Dataset}} & \multicolumn{12}{c|}{\textbf{Pathology}} \\ \hline
\multirow{2}{*}{\textbf{Train}} &  \multicolumn{4}{c|}{\textbf{STACH}} & \multicolumn{4}{c|}{\textbf{SBRAD}} & \multicolumn{4}{c|}{\textbf{PVC}} \\
 & \textbf{Sens.} & \textbf{Spec.} & \textbf{G-mean} & \multicolumn{1}{c|}{\textbf{F$_{2}$-score}} & \textbf{Sens.} & \textbf{Spec.} & \textbf{G-mean} & \multicolumn{1}{c|}{\textbf{F$_{2}$-score}} & \textbf{Sens.} & \textbf{Spec.} & \textbf{G-mean} & \textbf{F$_{2}$-score} \\ \hline
PTBXL\_rn50 & 0.951 & 0.964 & 0.958 & \multicolumn{1}{c|}{0.810} & 0.813 & 0.909 & 0.860 & \multicolumn{1}{c|}{0.521} & 0.956 & 0.978 & 0.967 & \multicolumn{1}{c|}{0.892} \\
TIS\_rn50 & \textbf{0.982} & 0.908 & 0.944 & \multicolumn{1}{c|}{0.669} & \textbf{0.961} & 0.849 & \textbf{0.903} & \multicolumn{1}{c|}{0.484} & \textbf{0.991} & 0.971 & 0.981 & \multicolumn{1}{c|}{0.899} \\
TIS\_tuned\_rn50 & 0.957 & \textbf{0.983} & \textbf{0.970} & \multicolumn{1}{c|}{\textbf{0.888}} & 0.836 & \textbf{0.921} & 0.877 & \multicolumn{1}{c|}{\textbf{0.561}} & 0.987 & \textbf{0.980} & \textbf{0.983} & \multicolumn{1}{c|}{\textbf{0.921}} \\ \hline
\end{tabular}
}
    \begin{tablenotes}
        \raggedright
        \item[] The best scores are represented by bold values.
    \end{tablenotes}
\end{threeparttable}

}
\end{table*}

\begin{table*}[htbp]
\caption{Comparison of the metrics of Inception networks: trained either on PTB-XL (PTBXL\_in) or TIS (TIS\_in) to detect one of the 7 abnormalities. DNNs are tested on TIS test data.} 
\label{table:Inception_1}
\resizebox{\textwidth}{!}{
\begin{threeparttable}
    \centering
    \begin{tabular}{|c|cccccccccccccccc|}
        \hline
        \multicolumn{1}{|c|}{\textbf{Dataset}} & \multicolumn{16}{c|}{\textbf{Pathology}} \\ \hline
\multirow{2}{*}{\textbf{Train}} &  \multicolumn{4}{c|}{\textbf{AFIB}} & \multicolumn{4}{c|}{\textbf{1AVB}} & \multicolumn{4}{c|}{\textbf{RBBB}} & \multicolumn{4}{c|}{\textbf{LBBB}} \\
 &  \textbf{Sens.} & \textbf{Spec.} & \textbf{G-mean} & \multicolumn{1}{c|}{\textbf{F$_{2}$-score}} & \textbf{Sens.} & \textbf{Spec.} & \textbf{G-mean} & \multicolumn{1}{c|}{\textbf{F$_{2}$-score}} & \textbf{Sens.} & \textbf{Spec.} & \textbf{G-mean} & \multicolumn{1}{c|}{\textbf{F$_{2}$-score}} & \textbf{Sens.} & \textbf{Spec.} & \textbf{G-mean} & \multicolumn{1}{c|}{\textbf{F$_{2}$-score}} \\ \hline
PTBXL\_in &   0.956 & 0.980 & 0.968 & \multicolumn{1}{c|}{0.831} & 0.824 & 0.907 & 0.864 & \multicolumn{1}{c|}{0.382} & 0.860 & \textbf{0.928} & 0.893 & \multicolumn{1}{c|}{0.706} & 0.793 & \textbf{0.982} & 0.882 & \multicolumn{1}{c|}{\textbf{0.559}} \\
TIS\_in &  \textbf{0.982} & \textbf{0.992} & \textbf{0.987} & \multicolumn{1}{c|}{\textbf{0.924}} & \textbf{0.963} & \textbf{0.927} & \textbf{0.945} & \multicolumn{1}{c|}{\textbf{0.500}} & \textbf{0.963} & 0.905 & \textbf{0.934} & \multicolumn{1}{c|}{\textbf{0.731}} & \textbf{0.911} & 0.966 & \textbf{0.938} & \multicolumn{1}{c|}{0.490} \\ \hline
\multicolumn{2}{c}{} &
    \end{tabular} 
\centering
\resizebox{\textwidth}{!}{
\begin{tabular}{|c|cccccccccccc|}
\hline
\multicolumn{1}{|c|}{\textbf{Dataset}} & \multicolumn{12}{c|}{\textbf{Pathology}} \\ \hline
\multirow{2}{*}{\textbf{Train}} & \multicolumn{4}{c|}{\textbf{STACH}} & \multicolumn{4}{c|}{\textbf{SBRAD}} & \multicolumn{4}{c|}{\textbf{PVC}} \\
 &  \textbf{Sens.} & \textbf{Spec.} & \textbf{G-mean} & \multicolumn{1}{c|}{\textbf{F$_{2}$-score}} & \textbf{Sens.} & \textbf{Spec.} & \textbf{G-mean} & \multicolumn{1}{c|}{\textbf{F$_{2}$-score}} & \textbf{Sens.} & \textbf{Spec.} & \textbf{G-mean} & \multicolumn{1}{c|}{\textbf{F$_{2}$-score}} \\ \hline
PTBXL\_in & 0.843 & 0.884 & 0.863 & \multicolumn{1}{c|}{0.670} & 0.529 & 0.915 & 0.696 & \multicolumn{1}{c|}{0.416} & 0.928 & 0.973 & 0.950 & \multicolumn{1}{c|}{0.737} \\
TIS\_in & \textbf{0.964} & \textbf{0.931} & \textbf{0.947} & \multicolumn{1}{c|}{\textbf{0.828}} & \textbf{0.975} & \textbf{0.933} & \textbf{0.954} & \multicolumn{1}{c|}{\textbf{0.760}} & \textbf{0.993} & \textbf{0.984} & \textbf{0.989} & \multicolumn{1}{c|}{\textbf{0.857}} \\ \hline
\end{tabular}
}
    \begin{tablenotes}
        \raggedright
        \item[] The best scores are represented by bold values.
    \end{tablenotes}
\end{threeparttable}

}
\end{table*}

\begin{table*}[htbp]
\caption{Comparison of the metrics of three Inception networks: pre-trained on TIS and fine-tuned on PTB-XL (TIS\_tuned\_in), and trained either on PTB-XL (PTBXL\_in) or TIS (TIS\_in) to detect one of the 7 abnormalities. DNNs are tested on PTB-XL test data.}
\label{table:Inception_2}
\resizebox{\textwidth}{!}{
\begin{threeparttable}
    \centering
    \begin{tabular}{|c|cccccccccccccccc|}
        \hline
        \multicolumn{1}{|c|}{\textbf{Dataset}} & \multicolumn{16}{c|}{\textbf{Pathology}} \\ \hline
\multirow{2}{*}{\textbf{Train}} &  \multicolumn{4}{c|}{\textbf{AFIB}} & \multicolumn{4}{c|}{\textbf{1AVB}} & \multicolumn{4}{c|}{\textbf{RBBB}} & \multicolumn{4}{c|}{\textbf{LBBB}} \\
 &  \textbf{Sens.} & \textbf{Spec.} & \textbf{G-mean} & \multicolumn{1}{c|}{\textbf{F$_{2}$-score}} & \textbf{Sens.} & \textbf{Spec.} & \textbf{G-mean} & \multicolumn{1}{c|}{\textbf{F$_{2}$-score}} & \textbf{Sens.} & \textbf{Spec.} & \textbf{G-mean} & \multicolumn{1}{c|}{\textbf{F$_{2}$-score}} & \textbf{Sens.} & \textbf{Spec.} & \textbf{G-mean} & \textbf{F$_{2}$-score} \\ \hline
PTBXL\_in& 0.942 & 0.964 & 0.953 & \multicolumn{1}{c|}{0.866} & 0.937 & 0.874 & 0.905 & \multicolumn{1}{c|}{0.566} & 0.951 & \textbf{0.953} & 0.952 & \multicolumn{1}{c|}{0.860} & 0.917 & 0.964 & 0.940 & \multicolumn{1}{c|}{0.739} \\
TIS\_in &  \textbf{0.956} & 0.952 & 0.954 & \multicolumn{1}{c|}{0.852} & 0.937 & 0.839 & 0.886 & \multicolumn{1}{c|}{0.508} & \textbf{0.994} & 0.926 & 0.959 & \multicolumn{1}{c|}{0.841} & \textbf{0.975} & 0.910 & \textbf{0.942} & \multicolumn{1}{c|}{0.598}\\
TIS\_tuned\_in & 0.935 & \textbf{0.980} & \textbf{0.957} & \multicolumn{1}{c|}{\textbf{0.896}} & \textbf{0.949} & \textbf{0.881} & \textbf{0.914} & \multicolumn{1}{c|}{\textbf{0.585}} & 0.988 & 0.942 & \textbf{0.965} & \multicolumn{1}{c|}{\textbf{0.866}} & 0.858 & \textbf{0.992} & 0.923 & \multicolumn{1}{c|}{\textbf{0.836}} \\ \hline
\multicolumn{2}{c}{} &
    \end{tabular} 
\centering
\resizebox{\textwidth}{!}{
\begin{tabular}{|c|cccccccccccc|}
\hline
\multicolumn{1}{|c|}{\textbf{Dataset}} & \multicolumn{12}{c|}{\textbf{Pathology}} \\ \hline
\multirow{2}{*}{\textbf{Train}} &  \multicolumn{4}{c|}{\textbf{STACH}} & \multicolumn{4}{c|}{\textbf{SBRAD}} & \multicolumn{4}{c|}{\textbf{PVC}} \\
 & \textbf{Sens.} & \textbf{Spec.} & \textbf{G-mean} & \multicolumn{1}{c|}{\textbf{F$_{2}$-score}} & \textbf{Sens.} & \textbf{Spec.} & \textbf{G-mean} & \multicolumn{1}{c|}{\textbf{F$_{2}$-score}} & \textbf{Sens.} & \textbf{Spec.} & \textbf{G-mean} & \textbf{F$_{2}$-score} \\ \hline
PTBXL\_in & 0.988 & 0.877 & 0.936 & \multicolumn{1}{c|}{0.628} & 0.789 & \textbf{0.920} & 0.852 & \multicolumn{1}{c|}{\textbf{0.532}} & 0.965 & 0.972 & 0.968 & \multicolumn{1}{c|}{0.880} \\
TIS\_in & \textbf{0.994} & 0.903 & 0.947 & \multicolumn{1}{c|}{0.665} & \textbf{0.984} & 0.836 & \textbf{0.907} & \multicolumn{1}{c|}{0.473} & \textbf{0.991} & 0.965 & 0.978 & \multicolumn{1}{c|}{0.881} \\
TIS\_tuned\_in & 0.988 & \textbf{0.915} & \textbf{0.951} & \multicolumn{1}{c|}{\textbf{0.690}} & 0.898 & 0.892 & 0.895 & \multicolumn{1}{c|}{0.530} & 0.987 & \textbf{0.979} & \textbf{0.983} & \multicolumn{1}{c|}{\textbf{0.918}} \\ \hline
\end{tabular}
}
    \begin{tablenotes}
        \raggedright
        \item[] The best scores are represented by bold values.
    \end{tablenotes}
\end{threeparttable}

}
\end{table*}

\begin{table*}[htbp]
\caption{Comparison of the metrics of LSTM networks: trained either on PTB-XL (PTBXL\_lstm) or TIS (TIS\_lstm) to detect one of the 7 abnormalities. DNNs are tested on TIS test data.}
    \label{table:LSTM_1}
\resizebox{\textwidth}{!}{
\begin{threeparttable}
    \centering
    \begin{tabular}{|c|cccccccccccccccc|}
        \hline
        \multicolumn{1}{|c|}{\textbf{Dataset}} & \multicolumn{16}{c|}{\textbf{Pathology}} \\ \hline
\multirow{2}{*}{\textbf{Train}} &  \multicolumn{4}{c|}{\textbf{AFIB}} & \multicolumn{4}{c|}{\textbf{1AVB}} & \multicolumn{4}{c|}{\textbf{RBBB}} & \multicolumn{4}{c|}{\textbf{LBBB}} \\
 &  \textbf{Sens.} & \textbf{Spec.} & \textbf{G-mean} & \multicolumn{1}{c|}{\textbf{F$_{2}$-score}} & \textbf{Sens.} & \textbf{Spec.} & \textbf{G-mean} & \multicolumn{1}{c|}{\textbf{F$_{2}$-score}} & \textbf{Sens.} & \textbf{Spec.} & \textbf{G-mean} & \multicolumn{1}{c|}{\textbf{F$_{2}$-score}} & \textbf{Sens.} & \textbf{Spec.} & \textbf{G-mean} & \multicolumn{1}{c|}{\textbf{F$_{2}$-score}} \\ \hline
PTBXL\_lstm & 0.863 & 0.829 & 0.846 & \multicolumn{1}{c|}{0.373} & 0.430 & 0.655 & 0.531 & \multicolumn{1}{c|}{0.081} & 0.899 & 0.895 & 0.897 & \multicolumn{1}{c|}{0.671} & 0.756 & \textbf{0.985} & 0.863 & \multicolumn{1}{c|}{\textbf{0.562}} \\
TIS\_lstm & \textbf{0.956} & \textbf{0.981} & \textbf{0.968} & \multicolumn{1}{c|}{\textbf{0.837}} & \textbf{0.922} & \textbf{0.912} & \textbf{0.917} & \multicolumn{1}{c|}{\textbf{0.436}} & \textbf{0.943} & \textbf{0.906} & \textbf{0.924} & \multicolumn{1}{c|}{\textbf{0.719}} & \textbf{0.902} & 0.963 & \textbf{0.932} & \multicolumn{1}{c|}{0.465} \\ \hline
\multicolumn{2}{c}{} &
    \end{tabular} 
\centering
\resizebox{\textwidth}{!}{
\begin{tabular}{|c|cccccccccccc|}
\hline
\multicolumn{1}{|c|}{\textbf{Dataset}} & \multicolumn{12}{c|}{\textbf{Pathology}} \\ \hline
\multirow{2}{*}{\textbf{Train}} & \multicolumn{4}{c|}{\textbf{STACH}} & \multicolumn{4}{c|}{\textbf{SBRAD}} & \multicolumn{4}{c|}{\textbf{PVC}} \\
 &  \textbf{Sens.} & \textbf{Spec.} & \textbf{G-mean} & \multicolumn{1}{c|}{\textbf{F$_{2}$-score}} & \textbf{Sens.} & \textbf{Spec.} & \textbf{G-mean} & \multicolumn{1}{c|}{\textbf{F$_{2}$-score}} & \textbf{Sens.} & \textbf{Spec.} & \textbf{G-mean} & \multicolumn{1}{c|}{\textbf{F$_{2}$-score}} \\ \hline
PTBXL\_lstm & 0.738 & 0.627 & 0.680 & \multicolumn{1}{c|}{0.393} & 0.257 & 0.884 & 0.476 & \multicolumn{1}{c|}{0.190} & 0.920 & \textbf{0.978} & 0.948 & \multicolumn{1}{c|}{0.761} \\
TIS\_lstm & \textbf{0.957} & \textbf{0.920} & \textbf{0.939} & \multicolumn{1}{c|}{\textbf{0.804}} & \textbf{0.980} & \textbf{0.941} & \textbf{0.960} & \multicolumn{1}{c|}{\textbf{0.785}} & \textbf{0.971} & 0.974 & \textbf{0.972} \\ \hline
\end{tabular}
}
    \begin{tablenotes}
        \raggedright
        \item[] The best scores are represented by bold values.
    \end{tablenotes}
\end{threeparttable}

}
\end{table*}

\begin{table*}[htbp]
\caption{Comparison of the metrics of three LSTM networks: pre-trained on TIS and fine-tuned on PTB-XL (TIS\_tuned\_lstm), and trained either on PTB-XL (PTBXL\_lstm) or TIS (TIS\_lstm) to detect one of the 7 abnormalities. DNNs are tested on PTB-XL test data.}
\label{table:LSTM_2}
\resizebox{\textwidth}{!}{
\begin{threeparttable}
    \centering
    \begin{tabular}{|c|cccccccccccccccc|}
        \hline
        \multicolumn{1}{|c|}{\textbf{Dataset}} & \multicolumn{16}{c|}{\textbf{Pathology}} \\ \hline
\multirow{2}{*}{\textbf{Train}} &  \multicolumn{4}{c|}{\textbf{AFIB}} & \multicolumn{4}{c|}{\textbf{1AVB}} & \multicolumn{4}{c|}{\textbf{RBBB}} & \multicolumn{4}{c|}{\textbf{LBBB}} \\
 &  \textbf{Sens.} & \textbf{Spec.} & \textbf{G-mean} & \multicolumn{1}{c|}{\textbf{F$_{2}$-score}} & \textbf{Sens.} & \textbf{Spec.} & \textbf{G-mean} & \multicolumn{1}{c|}{\textbf{F$_{2}$-score}} & \textbf{Sens.} & \textbf{Spec.} & \textbf{G-mean} & \multicolumn{1}{c|}{\textbf{F$_{2}$-score}} & \textbf{Sens.} & \textbf{Spec.} & \textbf{G-mean} & \textbf{F$_{2}$-score} \\ \hline
PTBXL\_lstm & 0.833 & 0.885 & 0.859 & \multicolumn{1}{c|}{0.649} & 0.506 & 0.735 & 0.610 & \multicolumn{1}{c|}{0.220} & 0.954 & 0.929 & 0.941 & \multicolumn{1}{c|}{0.818} & 0.908 & \textbf{0.964} & 0.936 & \multicolumn{1}{c|}{\textbf{0.735}} \\
TIS\_lstm & \textbf{0.946} & \textbf{0.942} & \textbf{0.944} & \multicolumn{1}{c|}{\textbf{0.822}} & \textbf{0.930} & 0.811 & 0.869 & \multicolumn{1}{c|}{0.468} & \textbf{0.988} & 0.925 & 0.956 & \multicolumn{1}{c|}{0.835} & \textbf{0.967} & 0.883 & 0.924 & \multicolumn{1}{c|}{0.532}\\
TIS\_tuned\_lstm & 0.908 & 0.939 & 0.923 & \multicolumn{1}{c|}{0.790} & 0.918 & \textbf{0.882} & \textbf{0.899} & \multicolumn{1}{c|}{\textbf{0.570}} & 0.972 &\textbf{0.952} & \textbf{0.962} & \multicolumn{1}{c|}{\textbf{0.873}} & 0.925 & 0.961 & \textbf{0.943} & \multicolumn{1}{c|}{\textbf{0.735}} \\ \hline
\multicolumn{2}{c}{} &
    \end{tabular} 
\centering
\resizebox{\textwidth}{!}{
\begin{tabular}{|c|cccccccccccc|}
\hline
\multicolumn{1}{|c|}{\textbf{Dataset}} & \multicolumn{12}{c|}{\textbf{Pathology}} \\ \hline
\multirow{2}{*}{\textbf{Train}} &  \multicolumn{4}{c|}{\textbf{STACH}} & \multicolumn{4}{c|}{\textbf{SBRAD}} & \multicolumn{4}{c|}{\textbf{PVC}} \\
 & \textbf{Sens.} & \textbf{Spec.} & \textbf{G-mean} & \multicolumn{1}{c|}{\textbf{F$_{2}$-score}} & \textbf{Sens.} & \textbf{Spec.} & \textbf{G-mean} & \multicolumn{1}{c|}{\textbf{F$_{2}$-score}} & \textbf{Sens.} & \textbf{Spec.} & \textbf{G-mean} & \textbf{F$_{2}$-score} \\ \hline
PTBXL\_lstm & 0.805 & 0.786 & 0.795 & \multicolumn{1}{c|}{0.392} & 0.570 & 0.762 & 0.659 & \multicolumn{1}{c|}{0.229} & 0.947 & \textbf{0.968} & 0.958 & \multicolumn{1}{c|}{\textbf{0.857}} \\
TIS\_lstm & \textbf{0.988} & 0.874 & 0.929 & \multicolumn{1}{c|}{0.602} & \textbf{0.969} & 0.845 & \textbf{0.905} & \multicolumn{1}{c|}{0.481} & 0.978 & 0.955 & 0.966 & \multicolumn{1}{c|}{0.843} \\
TIS\_tuned\_lstm & 0.951 & \textbf{0.953} & \textbf{0.952} & \multicolumn{1}{c|}{\textbf{0.775}} & 0.867 & \textbf{0.922} & 0.894 & \multicolumn{1}{c|}{\textbf{0.582}} & \textbf{0.987} & 0.953 & \textbf{0.970} & \multicolumn{1}{c|}{0.845} \\ \hline
\end{tabular}
}
    \begin{tablenotes}
        \raggedright
        \item[] The best scores are represented by bold values.
    \end{tablenotes}
\end{threeparttable}

}
\end{table*}

\balance
\twocolumn
\section*{Acknowledgment}

We thank Professor Andrei Tchernykh of CICESE Research Center, Mexico, for providing valuable feedback on the manuscript. We also thank Professor Philipp Kopylov of I.M. Sechenov First Moscow State Medical University for providing valuable assistance with medical inquiries.

\end{document}